\documentclass[reprint, amsmath, aip]{revtex4-1}
\usepackage{bm}
\usepackage{graphicx}

\newcommand*{\M}[0]{\ensuremath{\mathbf{M}}}
\newcommand*{\HOe}[0]{\ensuremath{\mathbf{H}_\text{Oe}}}

\begin{document}

\title{Nanosecond magnetization dynamics during spin Hall switching of in-plane magnetic tunnel junctions}

\author{Graham E. Rowlands\normalfont\textsuperscript{$\dagger$}}
\email{graham.rowlands@raytheon.com}
\altaffiliation[Now at: ]{Raytheon BBN Technologies, Cambridge, MA 02138}
\affiliation{Cornell University, Ithaca, New York 14853, USA}
\author{Sriharsha V. Aradhya\normalfont\textsuperscript{$\dagger$}}
\author{Shengjie Shi}
\author{Erin H. Yandel}
\author{Junseok Oh}
\affiliation{Cornell University, Ithaca, New York 14853, USA}
\author{Daniel C. Ralph}
\affiliation{Cornell University, Ithaca, New York 14853, USA}
\affiliation{Kavli Institute at Cornell, Ithaca, New York 14853, USA \protect\\ $\dagger$These authors contributed equally to this work.}
\author{Robert A. Buhrman}
\affiliation{Cornell University, Ithaca, New York 14853, USA}

\date{\today}
\revised{\today}

\begin{abstract}
We present a study of the magnetic dynamics associated with nanosecond scale magnetic switching driven by the spin Hall effect in 3-terminal nanoscale magnetic tunnel junctions (3T-MTJs) with in-plane magnetization. Utilizing fast pulse measurements in a variety of material stacks and detailed micromagnetic simulations, we demonstrate that this unexpectedly fast and reliable magnetic reversal is facilitated by the self-generated Oersted field, and the short-pulse energy efficiency can be substantially enhanced by spatial non-uniformity in the initial magnetization of the magnetic free layer.  The sign of the Oersted field is essential for this enhancement --- in simulations in which we artificially impose a field-like torque with a sign opposite to the effect of the Oersted field, the result is a much slower and stochastic switching process that is reminiscent of the so-called incubation delay in conventional 2-terminal spin-torque-switched MTJs.  
\end{abstract}

\maketitle

Spin transfer torque (STT) switching\cite{Slonczewski1996, Berger1996} of magnetic nanostructures has been extensively explored due to the potential for high performance magnetic memory technologies. Historically, in-plane-magnetized (IPM) all-metal spin valves based on the giant magnetoresistance effect were the first to be studied\cite{Katine2000}, and were observed to be switchable with current pulses as short as a few hundred picoseconds.\cite{Krivorotov2005, Devolder2005, Acremann2006} With the discovery of high tunneling magnetoresistance (TMR) with MgO tunnel barriers, magnetic tunnel junctions (MTJs) have become the primary focus of research and development, initially with magnetic layers having in-plane (IP) magnetic anisotropy and then with perpendicularly magnetized (PM) devices.\cite{Kubota2008, Sankey2008, Tomita2008, Devolder2008,Cui2010,Min2010, Lee2012, Rizzo2013} However, with IP MTJs it has proved difficult to achieve fast, reliable, and deterministic STT switching. Even as applied voltages approach the dielectric breakdown threshold,\cite{Schafers2009} the write error rates (WERs) remain unacceptably high in response to nanosecond-scale pulses.\cite{Min2010, Lee2012, Rizzo2013} This difficulty is often characterized in terms of ``incubation delay'' in both the thermally activated\cite{Cui2010} and the fast pulse switching regimes.\cite{Tomita2008, Devolder2008} This phenomenon has been variously ascribed to a bias-dependent field-like torque in MTJs,\cite{Garzon2009} to current and voltage feedback fluctuations in the MTJs during the write process,\cite{Garzon2009b} or to undesirable higher-order spin wave excitations in the magnetic free layer (FL).\cite{Devolder2008, Aurelio2010} The incubation delay in STT-switched IPM MTJs has motivated several alternate designs for short-pulse-switched MRAM.\cite{Liu2010, Rowlands2011, Luqiao2012b, Park2013, Park2014, Cubukcu2014}

\begin{figure}
\includegraphics[width=\linewidth]{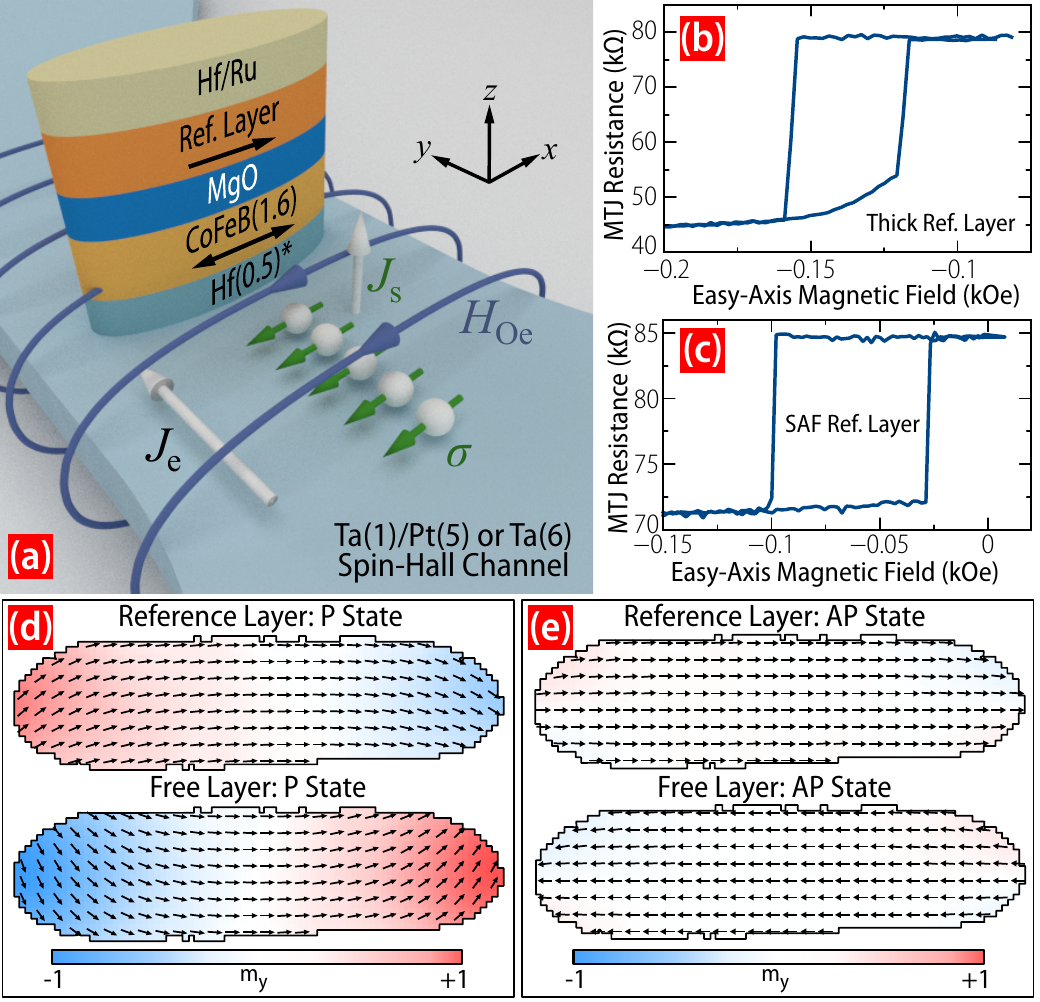}
\caption{\label{fig:1} (a) Schematic of the device with directions of charge current $J_e$, spin current $J_s$, spin accumulation $\sigma$, and Oersted field $\HOe$ indicated. The Hf insertion layer is only present in conjunction with the Ta/Pt/Hf channel. Easy axis minor hysteresis loops of the FL for devices with (b)  a thick RL and (c) an SAF RL. Simulated equilibrium configurations for the (d) P and (e) AP for the case of the thick RL.}
\end{figure}

\begin{figure}
\includegraphics[width=\linewidth]{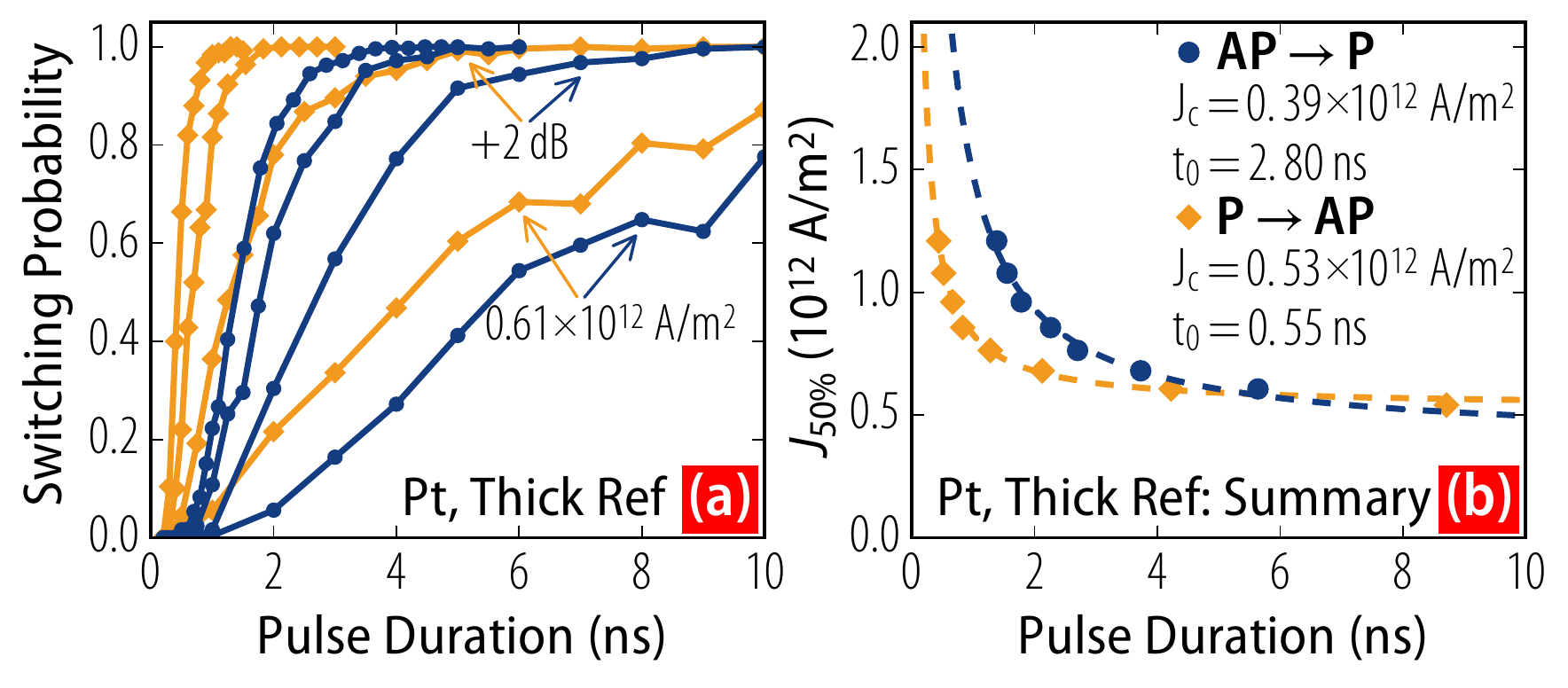}
\caption{\label{fig:2} (a) (orange diamonds) AP$\rightarrow$P and (blue circles) P$\rightarrow$AP switching probabilities vs. pulse length $\tau$ for a Pt-(thick ref) device. Each trace is for a constant $J$, in increments of 2 dB. Each point is averaged over 200 switching attempts. (b) Summary plot for Pt-(thick ref) showing the $J_\text{50\%}$ that produces 50\% switching for a given $\tau$.}
\end{figure}

\begin{table*}
\caption{Pulse switching characteristics from fits of Eq. 1 to experiments. The two entries marked by asterisks are poorly fit by Eq. 1}
\begin{ruledtabular}
\begin{tabular}{l l c c l c r}
& &  \multicolumn{2}{c}{AP$\rightarrow$P}  & \multicolumn{2}{c}{P$\rightarrow$AP}   \\
Channel & Reference Layer & $J_{c}$ ($10^{12}$ A/m$^2$) & $\tau_0$ (ns) &  $J_{c}$ ($10^{12}$ A/m$^2$) & $\tau_0$ (ns) & $H_\text{dip}$ \\
\hline
Ta	&      Thick&			0.47 $\pm$ 0.01&	0.75 $\pm$ 0.03&	0.47 $\pm$ 0.01&	0.91 $\pm$ 0.03 & 130 \\
Pt 	&	Thick&			0.39 $\pm$ 0.03&	2.8 $\pm$ 0.4	&	0.53 $\pm$ 0.01&	0.55 $\pm$ 0.02 &  200\\
Pt 	&	Pinned&			0.38 $\pm$ 0.02&	2.3 $\pm$ 0.2	&	0.41 $\pm$ 0.03 &	1.4 $\pm$ 0.2 & 60\\
Pt 	&	SAF &			0.40 $\pm$ 0.04&	1.8 $\pm$ 0.3	&	0.58 $\pm$ 0.04*& 	0.47 $\pm$ 0.07 & 30\\
Pt 	&	SAF+Weak Pinned&	0.26 $\pm$ 0.01&	2.9 $\pm$ 0.2	&	0.49 $\pm$ 0.04*& 	0.51 $\pm$ 0.07 & 55\\
Pt 	&	SAF+Strong Pinned&0.42 $\pm$ 0.04&	1.17 $\pm$ 0.19&	0.44 $\pm$ 0.05& 	1.18 $\pm$ 0.24 & 55
\end{tabular}
\end{ruledtabular}
\end{table*}

It is an important point, however, that incubation delays were not generally observed in the fast anti-damping (AD) STT switching regime of IPM all-metal spin valves,\cite{Krivorotov2005, Devolder2005, Acremann2006} clearly demonstrating that the pre-switching delay is neither an inherent feature of AD switching, nor is it purely related to thermal activation. The spin-Hall effect (SHE)\cite{Dyakonov1971a,Hirsch1999} provides an attractive alternate source of AD-STT for IPM devices. AD switching by the SHE utilizes a 3-terminal geometry (Fig. 1(a)): the transverse spin current arising from an applied IP current flowing in a heavy-metal channel can act to switch the magnetic FL of a MTJ patterned atop the channel, and the TMR of the MTJ is used to read out the FL orientation.\cite{Luqiao2012a} Charge current flow in the channel also produces an Oersted field ($\HOe$), but for channel materials of practical interest (e.g., Pt, $\beta$-Ta, and $\beta$-W), torque from the SHE acting on the FL dominates the switching dynamics and determines the sign of switching for a given current direction, irrespective of the sign of $\HOe$. This three-terminal magnetic tunnel junction (3T-MTJ) device approach (Fig. 1(a)) provides the opportunity for utilizing high amplitude short pulses to drive fast AD-STT reversal, as in spin-valve devices, while still incorporating a tunnel barrier with a high resistance area product (RA) to allow for fast readout.

While the SHE-induced reversal of IPM 3T-MTJs was first established in the thermally activated regime\cite{Luqiao2012a, Pai2012, Nguyen2015}, we have recently demonstrated surprisingly reliable switching of such devices on the nanosecond timescale (\textit{e.g}., $<10^{-5}$ write error rates with 2 ns pulses).\cite{Aradhya2016} Here we provide an in-depth analysis of the nanosecond-timescale magnetization dynamics that influence this highly desirable performance. First we demonstrate the robustness of fast switching behavior in 3T-MTJs with Pt and Ta spin Hall channels.\cite{Luqiao2012a,Nguyen2015} In some devices we observe an unexpected asymmetry, with substantial differences in the characteristic switching speeds between the antiparallel-to-parallel (AP$\rightarrow$P) and P$\rightarrow$AP switching polarities, and we explore this feature in detail using micromagnetic simulations. We find that this asymmetry stems from an interaction between the Oersted field $\HOe$ generated by current flow in the spin Hall channel and the micromagnetic non-uniformity present in the initial state of the FL due to dipole coupling with the magnetic reference layer (RL). We demonstrate through simulations that this mechanism can be beneficial: the Oersted field substantially increases switching speeds while an artificially reversed Oersted field does not. This stands in marked contrast to the detrimental effects of field-like torque observed in STT-switched two-terminal MTJs.\cite{Garzon2009} Finally we compare the fast-switching performance of Pt-based 3T-MTJs with RLs fabricated with varying pinning strengths and dipole field compensations, showing that very symmetric and fast AP$\rightarrow$P and P$\rightarrow$AP switching can be achieved with well-pinned synthetic antiferromagnetic (SAF) RLs.

\begin{figure}[t!]
\includegraphics[width=\linewidth]{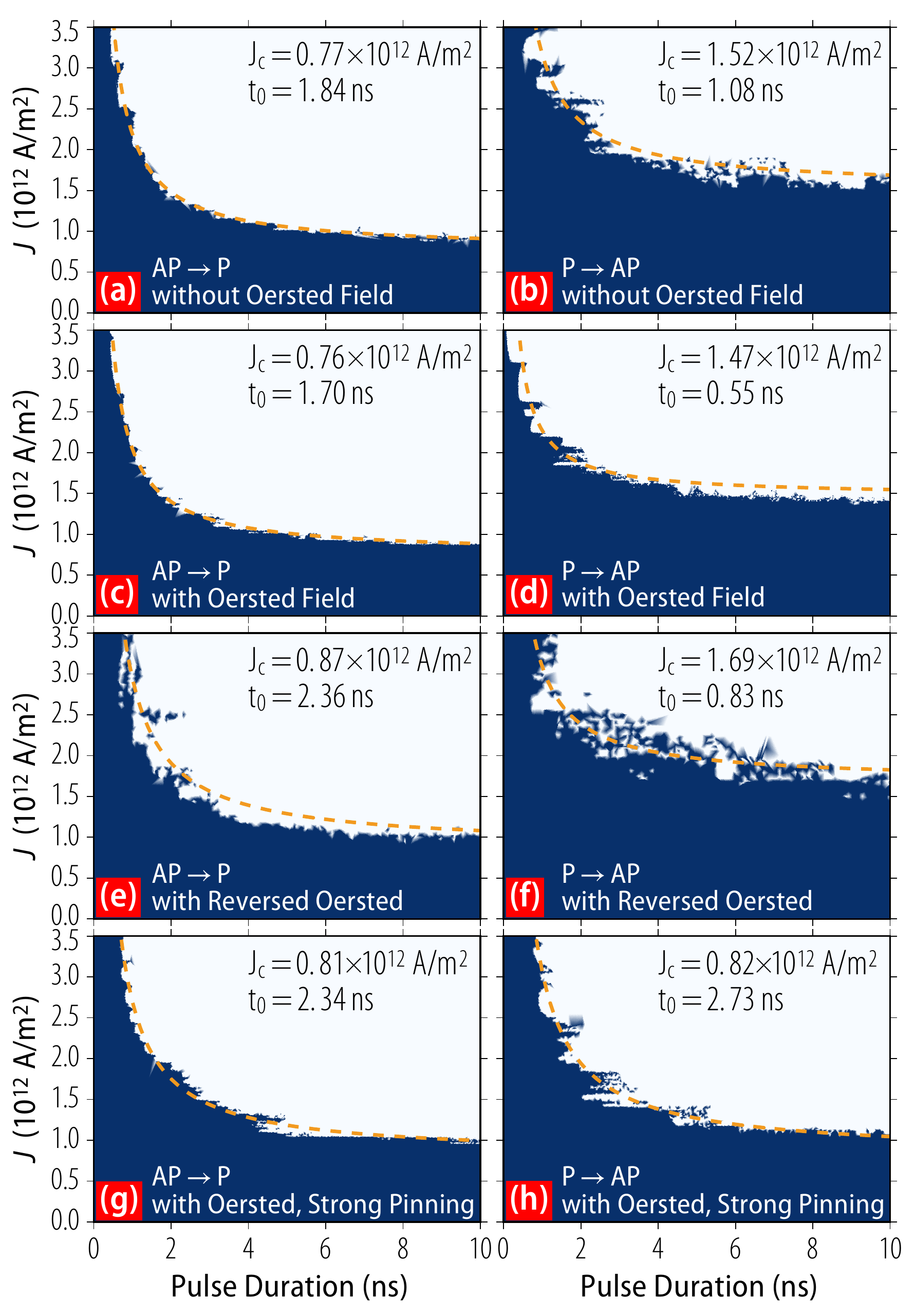}
\caption{Micromagnetic simulations of the switching phase diagrams of a Pt-(thick ref) and Pt-(pinned ref) device with strong pinning. (a,b) Pt-(thick ref) devices simulated without $\HOe$, (c,d) with $\HOe$ field, and (e,f) with an artificially inverted $\HOe$. (g,h) Pt-(pinned ref) device with a thinner 1.5 nm RL simulated with $\HOe$. The small spurs visible in, \textit{e.g.}, (a) and (c) are also present in macrospin simulations.\cite{supplement} Dashed lines show fits of Eq. 1 to the switching boundaries.}
\end{figure}

The devices analyzed in this paper were fabricated using the process introduced in our recent work,\cite{Aradhya2016} and the detailed stack structures for each type of device are presented in the supplementary information (SI).\cite{supplement}  Figure 1(b) shows the minor hysteresis loop of a Pt-based 3T-MTJ with a 4 nm thick FeCoB RL (hereafter referred to as a Pt-(thick ref) sample). The center of the loop is offset from zero field due to the stray field from the RL. The dipole interaction between the layers has additional important consequences. The rounding present in the P$\rightarrow$AP switching branch is a manifestation of dipole-induced rotation away from a uniform parallel configuration, as captured in simulations results shown in Fig. 1(d). Such rounding is greatly diminished in the AP$\rightarrow$P branch of the hysteresis loop since the dipole interaction in this case reinforces a uniform AP magnetization state of the two layers (Fig. 1(e)). The micromagnetic non-uniformity can also be reduced by using a SAF structure to both rigidly pin the RL magnetization and minimize its stray dipole field. The minor hysteresis loop of a Pt-SAF device is indeed much more square (Fig. 1(c)).

We now turn to the fast pulse switching measurements of Pt-(thick ref) and Ta-(thick ref) devices. Figure 2(a) shows the measured switching probability as a function of the pulse amplitudes and durations for a Pt-(thick ref) device in the P$\rightarrow$AP and AP$\rightarrow$P polarities. We summarize these data by extracting the interpolated current densities $J_\text{50\%}$ that result in 50\% switching probability for a given pulse duration $\tau$. These can then be fit to the macrospin model for anti-damping reversal,\cite{Sun2000} whereby 
\begin{equation}
J(\tau) = J_c ( 1 + \tau_0/\tau ),
\end{equation}
Here, $J_c$ is the critical switching voltage and $\tau_0$ the characteristic time at which switching is the most energy efficient. The $J_c$ and $\tau_0$ values from all studied devices are summarized in Table 1, while the data and fits to Eq. 1 for a Pt-(thick ref) device are shown in Fig. 2(b). We focus first on the large unexpected asymmetry in the switching timescales ($\tau_0$) between P$\rightarrow$AP and AP$\rightarrow$P polarities for this device. While it has been shown in simulations that non-ideal edge roughness profiles can alter the speed of reversal and introduce some polarity dependence into the switching speeds,\cite{Finocchio2007} this does not explain our data since we observe that the P$\rightarrow$AP switching is consistently faster than AP$\rightarrow$P across all Pt-(thick ref) devices we have measured. Furthermore, the opposite tendency is observed in Ta-(thick ref) devices (faster AP$\rightarrow$P switching), even though these have similar edge profiles. The size of the asymmetry is less pronounced in Ta-(thick ref) compared to Pt-(thick ref) devices. As discussed below, these observations can be related to the different signs and strengths of the SHE in Pt and Ta.

Motivated by the influence of the micromagnetic non-uniformity on magnetic-field driven P$\rightarrow$AP switching, we have used simulations to explore several scenarios for fast-pulse spin torque switching.\cite{Donahue1999} In each scenario we simulated the entire micromagnetic switching phase diagram at $T = 0$ K, including the full dynamic response of the RL (details in SI).\cite{supplement} First, we confirmed that the reversal characteristics of an isolated magnetic FL driven only by the SHE torque, not including the effects of $\HOe$ from the applied current, are independent of the current polarity (data not shown). Next, we introduced an unpinned 3.0 nm thick RL that couples to the FL via the dipolar interaction, (Figs. 3(a,b)). In this case, the characteristic $J_c$ and $\tau_0$ of the two polarities diverged: while the AP$\rightarrow$P transition proceeded in a relatively uniform manner the P$\rightarrow$AP transition required an increased $J_c$ and proceeded more slowly via complex micromagnetic dynamics that take place in both the FL and RL. These simulations are in conflict with the data of Fig. 2(b), where the P$\rightarrow$AP transition appears greatly accelerated. However, upon introducing $\HOe$ from the SHE channel in the simulations, we saw a dramatic reduction in $\tau_0$ for P$\rightarrow$AP switching, as well as a commensurate reduction of the fine structure (Figs. 3(c,d)). In the case of Pt 3T-MTJ devices, the sign of the spin Hall torque is such that (for either sign of current) $\HOe$ aids the switching driven by the SHE by helping to destabilize the initial state. Given the substantial P state micromagnetic non-uniformity, the out-of-plane $-\M \times \HOe$ torque acts to quickly increase the magnitude of pre-switching dynamics in P$\rightarrow$AP switching and thereby decrease their duration. On the other hand, the lack of an appreciable $-\M \times \HOe$ torque in the uniform AP state (where $\M || \HOe$) leaves the AP$\rightarrow$P transition largely unchanged with respect to $\tau_0$ and $J_c$, although inspection of individual transitions showed a more coherent reversal with $\HOe$ present. We verified that the beneficial effect of $\HOe$ occurs primarily during the early stages of the switching process --- it only needs to be present for the first 0.25 ns of the current pulse to facilitate fast P$\rightarrow$AP reversal.

When we tested the influence of $\HOe$ by un-physically reversing its sign in the simulations we obtained undesirable, highly non-uniform pre-switching dynamics that slow the switching, especially in the P$\rightarrow$AP polarity (Figs. 3(e,f)). This behavior is reminiscent of the incubation delay in STT switched 2-terminal MTJs.\cite{Tomita2008, Devolder2008, Cui2010, Garzon2009, Garzon2009b, Aurelio2010}, and also helps explain the experimental results with Ta-(thick ref) devices. The SHE in Ta is opposite in sign to Pt, so $\HOe$ in this case acts to stabilize the initial state, thus opposing the SHE. In Table 1 of the SI we see that AP$\rightarrow$P switching in Ta-(thick ref) devices is marginally faster (since here the P$\rightarrow$AP switching is slowed by $\HOe$). However the effect is small since the increased magnitude of $\xi_\text{SH} = -0.12$\cite{Luqiao2012a} compared to $\xi_\text{SH} = +0.08$ in Pt reduces the relative importance of HOe. Finally, we simulated a Pt device with a thinner (1.5 nm) but well-pinned RL ($H_\text{pin} = 2$ kOe), which resulted in a reduced dipolar field at the FL and a reduction of the P-state non-uniformity. In this case we obtained a much closer correspondence of $J_c$ and $\tau_0$ across the two polarities (Figs. 3(g,h)). However, $\tau_0$ was relatively long in both cases, $> 2$ ns, which we attribute to the strong coercive field ($\sim 200$ Oe) of the simulated FL and the lack of thermal effects. The simulations do show reliable sub-ns switching at sufficiently strong current pulses, though with a correspondingly reduced energy efficiency.

We conclude from these initial experimental results and simulations that micromagnetic non-uniformity in the FL of a 3-T MTJ device is conducive to enabling fast and efficient STT driven reversal, with $\tau_0 < 1$ ns, provided $\HOe$ generated by the current in the spin Hall channel is in the direction to encourage further deformation, \textit{i.e.}, opposite to the overall coercive field.  As the result, large amplitude oscillations are quickly established, which suppresses the formation of localized non-uniform magnetic modes across the FL and speed the overall reversal. When this initial non-uniformity is reduced, as in the AP state for a Pt-(thick ref) device, the reversal is slower although still fast and evidently devoid of an incubation delay.  

\begin{figure}
\includegraphics[width=\linewidth]{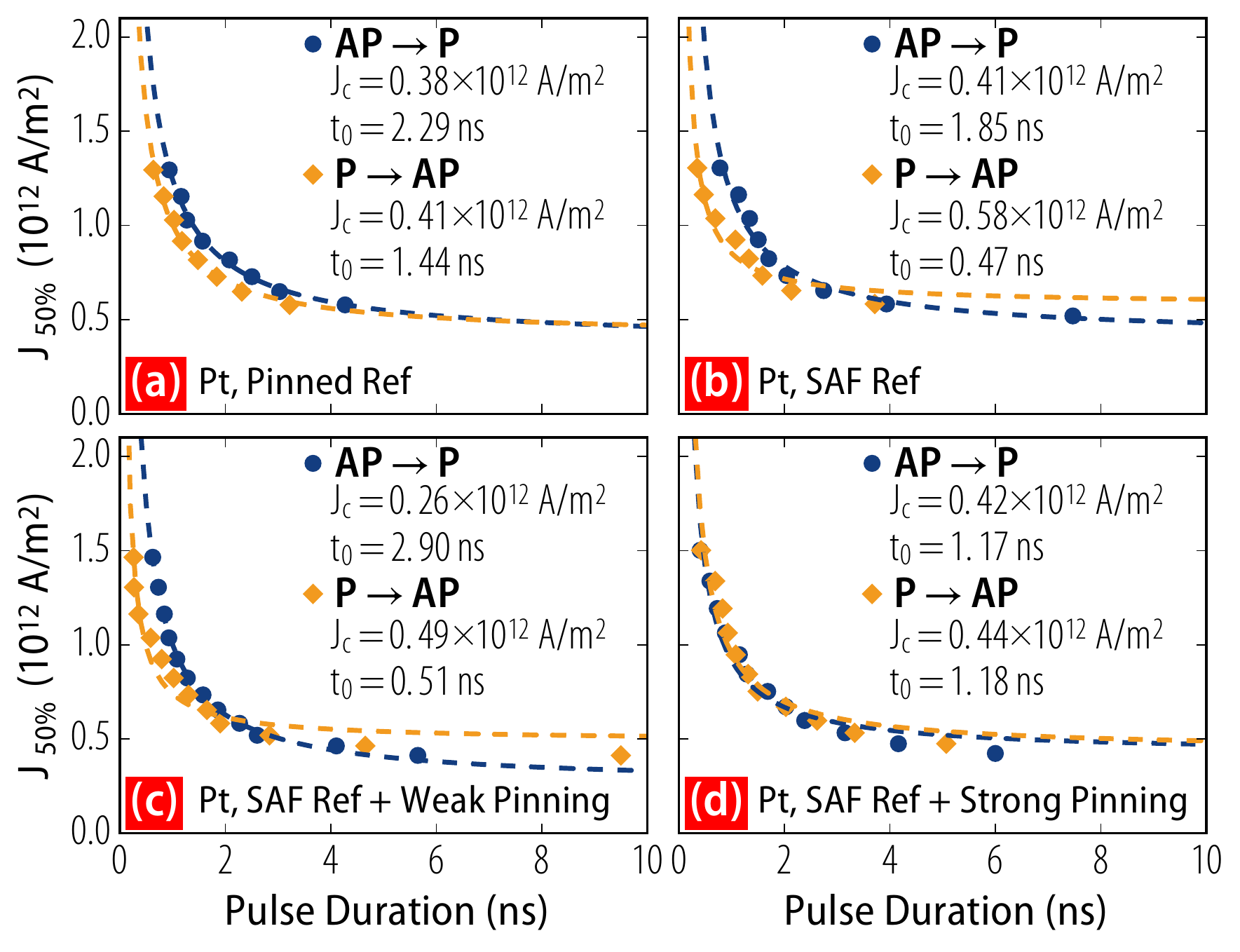}
\caption{Plots of 50\% switching behavior for (a) a Pt-(pinned ref) device, (b) a Pt-SAF device, (c) a Pt-(SAF+weak pinning) device, and (d) a Pt-(SAF+strong pinning) device.}
\end{figure}

Based on these conclusions, we expected that reducing dipole interactions between the magnetic layers and increasing the rigidity of the RL would reduce the asymmetry between the switching polarities, which is desirable in most applications, while still providing relatively fast switching in both directions. Therefore, we fabricated Pt-based 3T-MTJ devices with RLs of increasing complexity. Figures 4(a-d) show the switching summaries from four such device types (see SI for details): a) a Pt-(pinned ref) device with an exchange pinned RL, b) a Pt-SAF device with a SAF RL, and (c,d) two Pt-(SAF + pinning) devices with exchange pinned SAF RLs of two different pinning strengths. Compared to the Pt-(thick ref) device, the dipole field in the Pt-(pinned ref) sample was reduced by a factor of four, and in the Pt-SAF and Pt-(SAF + pinning) devices by an order of magnitude. While our SAF could be further fine-tuned, even if the average dipole moment were reduced to zero there would still be significant non-uniform dipolar coupling to the FL. We find that the Pt-(pinned ref) device yields a more symmetric $\tau_0$ compared to the Pt-(thick ref) sample, and can be well fit to the macrospin model of Eq. 1, implying more coherent switching dynamics (Fig. 4(a)). The Pt-SAF device also shows a more symmetric $\tau_0$, but cannot be well fit to Eq. 1 (Fig. 4b). This is likely due to dynamics originating in the thin, weakly pinned RL. In the two Pt-(SAF + pinning) devices, the strength of the pinning was varied by annealing the devices in different strengths of magnetic field. In the first case (the Pt-(SAF + weak pinning) device) we deliberately annealed the device in a low (0.15 T) field to weaken the exchange pinning\cite{Kerr2005} while still establishing a preferred direction for the RL (Fig. 4(c)). In the second case (Pt-(SAF + strong pinning)) we used a high (0.5 T) field to provide a stronger exchange pinning (Fig. 4(d)). The Pt-(SAF + weak pinning) device exhibits switching characteristics similar to those of the Pt-SAF sample, while the Pt-(SAF + strong pinning) sample shows a $\tau_0$ of approximately 1 ns in both polarities. Since the weakly pinned SAF still supports coupled RL/FL dynamics, we conclude that a very strong pinning is needed to prevent this detrimental behavior. Finally, in all cases we note that the switching timescales for P$\rightarrow$AP and AP$\rightarrow$P switching converge in response to longer pulses, even if the macrospin fits do not capture this fact. This is because the initial dynamics are less important over longer timescales. 

In conclusion, we have demonstrated robust ns-timescale reversal of 3T-MTJ devices utilizing spin Hall torque from both a Ta and Pt channel. Extremely fast switching, with a characteristic reversal time $\tau_0 \ll  1$ ns, can be obtained when there is substantial micromagnetic non-uniformity in the FL and the pulsed Oersted field is in the direction that encourages switching, making the formation of localized magnetic deformations within the FL less likely during reversal. If the direction of $\HOe$ is reversed in the simulations, switching takes substantially longer and becomes highly non-uniform. Since the FL micromagnetics vary substantially between the P and AP configuration in MTJs with weakly pinned thick RLs, the switching speeds in that case are quite different between P$\rightarrow$AP and AP$\rightarrow$P reversal. With reduced FL micromagnetic non-uniformity courtesy of a SAF RL, and strong pinning that minimizes RL oscillations, somewhat slower (but still fast) switching with $\sim 1$ ns is obtained since the $\HOe$ still promotes uniform and reliable reversal. The understanding of the fast switching dynamics reported here could provide opportunities for additional optimization of 3T-MTJs through explicit exploitation of micromagnetic effects. This would further enhance the attractiveness of 3T-MTJs, with their separation of low impedance write and high impedance read paths, for applications requiring fast switching.

The authors thank C. L. Jermain for help with FMR characterization, M.-H. Nguyen for help with the film stack development, and P. G. Gowtham and Y. Ou for stimulating discussions. The research is based upon work supported by the Office of the Director of National Intelligence (ODNI), Intelligence Advanced Research Projects Activity (IARPA), via contract W911NF-14-C0089. The views and conclusions contained herein are those of the authors and should not be interpreted as necessarily representing the official policies or endorsements, either expressed or implied, of the ODNI, IARPA, or the U.S. Government. The U.S. Government is authorized to reproduce and distribute reprints for Governmental purposes notwithstanding any copyright annotation thereon. Additionally, this work was supported by the NSF/MRSEC program (DMR-1120296) through the Cornell Center for Materials Research, by the Office of Naval Research, and by the NSF (Grant No. ECCS-0335765) through use of the Cornell Nanofabrication Facility/National Nanofabrication Infrastructure Network.

%

\end{document}